# Constructing LDPC Codes by $2$-Lifts

Xudong Ma and En-hui Yang
Dept. of Electrical and Computer Engineering
University of Waterloo
Waterloo, Ontario, Canada N2L 3G1
Email: {x3ma,ehyang}@bbcr.uwaterloo.ca

**Abstract**

We propose a new low-density parity-check code construction scheme based on 2-lifts. The proposed codes have an advantage of admitting efficient hardware implementations. With the motivation of designing codes with low error floors, we present an analysis of the low-weight stopping set distributions of the proposed codes. Based on this analysis, we propose design criteria for designing codes with low error floors. Numerical results show that the resulting codes have low error probabilities over binary erasure channels.

## I. Introduction

Low-Density Parity-Check (LDPC) codes receive much attention recently. The advantages of the codes include capacity-approaching performance for many important channels, and highly efficient parallel decoding algorithms.

The *protograph based* LDPC codes are a class of LDPC codes constructed by random $N$-lifts of small Tanner graphs [1]. A major advantage of protograph based LDPC codes is that decoding can be efficiently implemented by a semi-parallel hardware structure [2]. Several protograph based LDPC code ensembles with low error probabilities are also previously reported, for example, [3], [4], [5].

In this paper, we propose a new LDPC code construction scheme based on a series of random $2$-lifts. The proposed code ensembles are subsets of protograph based code ensembles by $2^n$-lifts, where $n$ is a positive integer. As with the protograph based codes, decoding of the proposed codes can be efficiently implemented by the same semi-parallel hardware structure. In addition, the description complexity of the proposed codes is lower than that for conventional protograph based codes.

It is well known that LDPC codes may exhibit error floors. The error floors are caused by small subgraphs with certain structures in the corresponding Tanner graphs [6], [7], [8], [9]. One of the above subgraph structures is the so called *stopping set*.

In order to understand and predict the performance of LDPC codes, previous research has proposed several approaches on approximately calculating stopping set distributions and codeword distributions [10], [11]. These previous approaches determine the exact values of limiting exponents by large deviation principles; in here, exponents are defined as the logarithms of the numbers of codewords or stopping sets with certain relative distances, normalized by the blocklengths.

For properly designed LDPC codes, the limiting exponents with respect to blocklengths are typically negative for low relative distances and zero for vanishing relative distances. In other words, low-weight stopping sets are typically stopping sets with vanishing relative distances; the number of these low-weight stopping sets scales sub-exponentially with respect to blocklength. These potentially existing low-weight stopping sets result in high error floors. It is desirable that these low-weight stopping sets are avoided.

In this paper, we present an analysis of low-weight stopping set distributions of the proposed codes in terms of graph expansion properties. We show that low-weight stopping sets can be largely avoided if the protograph satisfies certain graph expansion properties. Based on this analysis, we propose design criteria for the proposed $2$-lift based codes to have low error floors. Several numerical examples of code design using the design criteria are presented. The numerical results confirm our theoretical analysis.

The rest of this paper is organized as follows. In Section II, we introduce notations and backgrounds. The proposed code construction scheme is shown in Section III. In Section IV, we present the analysis on low-weight stopping set distributions. Remarks and discussions are presented in Section V. Based on the above discussions, a set of design criteria is presented. The proof is shown in Section VI. We present numerical results in Section VII. Conclusions are given in Section VIII.

## II. Notation and Preliminary

A Binary Erasure Channel (BEC) is shown in Fig. 1. BEC is a model for packet network communications, where main impairments are packet loss. The channel receives binary inputs and outputs $0$, $1$, or $e$. A transmitted signal is received correctly at the receiver with probability $1-\xi$. With probability $\xi$, the receiver receives an erasure symbol $e$, where the transmitted bit is completely lost. The channel conditional probabilities are summarized as follows:

$$\mathbb{P}(y=1|x=1)=\mathbb{P}(y=0|x=0)=1-\xi,$$

$$\mathbb{P}(y=0|x=1)=\mathbb{P}(y=1|x=0)=0,$$

$$\mathbb{P}(y=e|x=1)=\mathbb{P}(y=e|x=0)=\xi,$$

where, $x$ stands for the transmitted signal, and $y$ stands for the received signal.

For LDPC codes transmitted over a BEC, a transmitted codeword can not be recovered if and only if the set of erased bits contains a stopping set. A formal definition of stopping set is given as follows.

**Definition** A nonempty variable node set $\mathcal{S}$ is called a *stopping set*, if each check node in the Tanner graph is either connected to at least two variable nodes in $\mathcal{S}$, or is not connected to any variable node in $\mathcal{S}$.

We will use the following widely adopted graph theoretical notations throughout this paper. Given a graph $G$, we use the notation $G=(V,E)$ to denote that $V$ is the set of all vertices and $E$ is the set of all edges. We use $(u,v)$ to denote the undirected edge between two vertices $u$ and $v$. For a set of vertices $\mathcal{S}$ in a graph $G=(V,E)$, we denote the neighborhood of $\mathcal{S}$ by $N(\mathcal{S})$,

$$N(\mathcal{S})=\{u:(u,v)\in E, v\in\mathcal{S}\}.$$

We use $|\mathcal{S}|$ to denote the cardinality of a set $\mathcal{S}$. We say that a graph $G'=(V',E')$ is a subgraph of $G=(V,E)$ induced by $V'$ if $V'\subset V$ and $E'$ is the set of edges in $E$ joining two vertices in $V'$.

We will need the following graph expansion properties in our later discussions. Let $G$ denote a Tanner graph.

**Definition** We say that a variable node set $\mathcal{S}$ in $G$ has $\beta$ vertex expansion property, if $|N(\mathcal{S})|\geq\beta|\mathcal{S}|$.

**Definition** Let $\mathcal{S}$ denote a variable node set in $G$. Let $\widetilde{N}(B)$ denote the set of check nodes connected to nodes in a variable node set $B$ at least twice. We say that $\mathcal{S}$ has $(\gamma,\eta)$ uniform-edge-distribution property if for each subset $B\subset\mathcal{S}$ with $|B|\leq\eta|\mathcal{S}|$, we have

$$|\widetilde{N}(B)|\leq\gamma|B|. \tag{1}$$

We use $\lfloor\cdot\rfloor$ and $\lceil\cdot\rceil$ to denote the flooring function and the ceiling function respectively. That is, $\lfloor x\rfloor$ is the largest integer smaller than a real number $x$, and $\lceil x\rceil$ is the smallest integer greater than the real number $x$. We use $H_e(x)$ to denote the entropy function with base $e$,

$$H_e(x)=-x\log_e(x)-(1-x)\log_e(1-x)$$

where $0<x<1$. We use $H_2(x)$ to denote the entropy function with base $2$. We denote the set difference of two sets $A$, and $B$ as $A\backslash B$,

$$A\backslash B=\{u|u\in A, u\notin B\}.$$

Throughout this paper, we assume all graphs are simple graphs; that is, the graphs do not contain loop and parallel edges.

## III. Code Construction Scheme

In this Section, we present the proposed code construction scheme. We need the following definition for $N$-lift graphs. A similar definition can also be found in [1].

**Definition** Given a graph $G = (V, E)$, a graph $\widehat{G} = (\widehat{V}, \widehat{E})$ is said to be a $N$-lift of $G$ if the following conditions hold.

- The vertex set $\widehat{V}$ is the union of $N$ disjoint vertex sets $V_1, \ldots, V_N$, where $V_1, \ldots, V_N$ have the same cardinality as $V$. That is,

$$\widehat{V} = V_1 \cup \cdots \cup V_N,$$

$$|V| = |V_1| = \cdots = |V_N|.$$

- There exists a covering map $\pi : V_1 \times \cdots \times V_N \to V$.
- Given an edge $(u, v)$ in the graph $G$, if a vertex $u_1 \in \pi^{-1}(u) \subset \widehat{V}$, then $u_1$ is connected to exactly one vertex in $\pi^{-1}(v)$.

We call the graph $G$ the base graph of $\widehat{G}$, and the map $\pi(\cdot)$ the covering map from $\widehat{G}$ to $G$.

A $N$-lift can also be visualized as an operation of *copying $N$ times and edge-permutating*. Two examples of 2-lift graphs are shown in Fig. 2. The 2-lift graph shown in sub-figure (a) consists of two identical copies of the base graph without edge permutation. The 2-lift graph shown in sub-figure (b) consists of two copies of the base graph with one pair of edges permuted. In the sequel, we say that a graph $\widehat{G}$ is a uniformly random 2-lift of a graph $G$, if $\widehat{G}$ is randomly chosen as one of 2-lifts of $G$ with equal probability.

**The Proposed Code Construction Scheme:** A small sized Tanner graph is first constructed by computational searching or algebraic construction. We denote this Tanner graph by $G_0$ and call it the *protograph*. The Tanner graph of the resulting code is obtained by recursively applying the uniformly random 2-lift procedure several times. That is, we construct a sequence of Tanner graphs $G_1, \ldots, G_k, \ldots, G_M$, such that $G_k$ is a uniformly random 2-lift of $G_{k-1}$, for $k = 1, \ldots, M$. The graph $G_M$ is the Tanner graph of the resulting code.

## IV. Low-Weight Stopping Set Distribution

In this section, we will present an analysis for the distributions of low-weight stopping sets. We will need the following notations.

- As with Section III, we denote the sequence of Tanner graphs obtained in the code construction scheme by $G_0, G_1, \ldots, G_M$.
- Denote the covering map from $G_k$ to $G_{k-1}$ by $\pi_k(\cdot)$.
- Denote the number of variable nodes in the protograph by $n_0$.
- Denote the minimal weight of stopping sets in the protograph by $d_0$,

$$d_0 = \min_{\mathcal{S} \text{ is a stopping set}} |\mathcal{S}|. \tag{2}$$

*Proposition 4.1:* Let $G$, $\widehat{G}$ be two Tanner graphs, where the graph $\widehat{G}$ is a $N$-lift of $G$. Let $\pi$ denote the covering map from $\widehat{G}$ to $G$. If $\widehat{\mathcal{S}}$ is a stopping set in the $N$-lift $\widehat{G}$, then the set $\mathcal{S} = \pi(\widehat{\mathcal{S}})$ is a stopping set in the base graph $G$.

*Corollary 4.2:* In the proposed code construction scheme, the minimal weight of stopping sets in $G_M$ is greater than or equal to $d_0$.

**Condition A:** Let $\beta$, $\gamma$, $\eta$, $\overline{d}$ be real numbers, $0 < \beta$, $0 < \gamma < 1$, and $\overline{d} > 0$. We say that a set $\mathcal{S}$ in a Tanner graph $G$ satisfies Condition (A) with respect to $\beta$, $\gamma$, $\eta$, $\overline{d}$, if the following conditions hold.

- The set $\mathcal{S}$ have $\beta$ vertex expansion property.
- The set $\mathcal{S}$ have $(\gamma, \eta)$ uniform-edge-distribution property.
- The neighboring check nodes $N(\mathcal{S})$ of $\mathcal{S}$ in the graph $G$ have an average degree less than $\overline{d}$ in the subgraph induced by $\mathcal{S} \cup N(\mathcal{S})$.

In the sequel, we say that a variable node set have strong graph expansion properties, if the set satisfies the Condition (A) with a large $\beta$ and a small $\gamma$.

*Theorem 4.3:* Let $\mathcal{S}$ be a stopping set in the Tanner graph $G_k$. Assume that the set $\mathcal{S}$ satisfies Condition (A) with respect to $\beta$, $\gamma$, $\eta$, $\overline{d}$. Let $\theta$ be a real number such that $\theta|\mathcal{S}|$ is an integer, $\theta \geq 1-\eta$. Let $\mathfrak{B}(\mathcal{S}, \theta)$ denote the number of stopping sets $\widehat{\mathcal{S}}$ in $G_{k+1}$ with $|\widehat{\mathcal{S}}| = (1+\theta)|\mathcal{S}|$, $\pi_{k+1}(\widehat{\mathcal{S}}) = \mathcal{S}$. Then, the expectation $\mathbb{E}\left[\mathfrak{B}(\mathcal{S}, \theta)\right]$ is upper bounded as follows.

$$\mathbb{E}\left[\mathfrak{B}(\mathcal{S}, \theta)\right] \leq \exp\left\{\left[H_e(\theta) + (1-\theta)\log_e 2 - 2(\beta - \gamma\theta)\left(\frac{1}{2}\right)^{(\beta\overline{d}/(\beta-\gamma\theta))}\right]|\mathcal{S}|\right\}. \tag{3}$$

*Corollary 4.4:* Let $\theta_0$ be a positive real number such that, $\theta_0 \leq 1-\eta$, and for all $0 < \theta \leq \theta_0$,

$$H_e(\theta) + (1-\theta)\log_e 2 - 2(\beta - \gamma\theta)\left(\frac{1}{2}\right)^{(\beta\overline{d}/(\beta-\gamma\theta))} \leq -\Delta, \tag{4}$$

where $\Delta$ is a positive real number. Then, with high probability, there is no stopping set $\widehat{\mathcal{S}}$ with $\pi_k(\widehat{\mathcal{S}}) = \mathcal{S}$, $|\widehat{\mathcal{S}}| \leq (1+\theta_0)|\mathcal{S}|$ in $G_{k+1}$, if $|\mathcal{S}|$ is sufficiently large.

*Lemma 4.5:* Let $\mathcal{S}$ be a stopping set in the graph $G_k$. Assume that $\mathcal{S}$ satisfies Condition (A) with respect to $\beta$, $\gamma$, $\eta$, $\overline{d}$. Let $\widehat{\mathcal{S}}$ be a stopping set in the graph $G_{k+1}$ with $\pi_{k+1}(\widehat{\mathcal{S}}) = \mathcal{S}$, $|\widehat{\mathcal{S}}| = (1+\theta)|\mathcal{S}|$, $\theta \geq \theta_0$. Then, $\widehat{\mathcal{S}}$ satisfies Condition (A) with respect to $\beta'$, $\gamma'$, $\eta'$, $\overline{d'}$, where

$$\beta' \geq \min_{\theta_0 \leq \theta \leq 1} \frac{2\beta - \gamma(1-\theta)}{1+\theta}$$
$$\gamma' \leq \frac{3}{2}\gamma, \;\; \eta' = 2(1-\theta_0), \;\; \overline{d'} \leq \frac{\beta\overline{d}}{\beta'} \tag{5}$$

For the special case where $G_0, \ldots, G_k$ are $(d_v, d_c)$ regular LDPC codes, $\widehat{\mathcal{S}}$ satisfies Condition (A) with respect to $\beta'$, $\gamma'$, $\eta'$, $\overline{d'}$, where

$$\beta' \geq \min_{\theta_0 \leq \theta \leq 1} \frac{2\beta - \gamma(1-\theta)}{1+\theta}. \tag{6}$$

In this special case, the following relationships always hold:

$$\gamma' \leq \frac{d_v}{2}, \;\; \eta' = 0, \;\; \overline{d'} \leq \frac{d_v}{\beta'}. \tag{7}$$

## V. Remark and Discussion

*Remark 1:* Proposition 4.1 and Corollary 4.2 show that the minimal weight of stopping sets in the resulting Tanner graph is always not less than the minimal weight of stopping sets in the protograph. Hence, it is desirable that the protograph has a large minimal weight of stopping sets.

*Remark 2:* According to Theorem 4.3 and Corollary 4.4, with high probability, there is no stopping sets with weight less than $(1+\theta_0)|\mathcal{S}|$. In order to have a large $\theta_0$, $\overline{d}$ should be minimized and $\beta$, $\gamma$ should be maximized. The above discussion implies that stopping sets with strong graph expansion properties are less problematic in terms of introducing low-weight stopping sets. Hence, the protograph $G_0$ in the code construction scheme should be chosen so that stopping sets with weak graph expansion properties are avoided.

*Remark 3:* Lemma 4.5 implies that, with high probability, the stopping sets $\widehat{\mathcal{S}}$ with $\pi_{k+1}(\widehat{\mathcal{S}}) = \mathcal{S}$ also have strong graph expansion properties. Lemma 4.5 presents a worst-case analysis. We conjecture that $\widehat{\mathcal{S}}$ have much stronger graph expansion properties in terms of average case analysis.

*Remark 4:* In hardware implementations, Tanner graphs need to be stored. It can be checked that $O(m)$ bits are needed to store a Tanner graph of the proposed codes, where $m$ is the number of edges. For general random-constructed LDPC codes, $\log_2(m!) \approx O(m\log_2(m))$ bits are needed. The description complexity of the proposed codes is much lower.

Based on the above discussion, we propose the following design criteria for designing codes with large stopping distance:

- the protograph should have a large minimal weight of stopping sets;
- the protograph should not contain low-weight stopping sets with weak graph expansion properties.

## VI. Proof of Theorem 4.3

We first consider a randomly generated variable node set $\widehat{\mathcal{S}}$ in the graph $G_{k+1}$ chosen from all variable node sets $A$, $\pi_{k+1}(A) = \mathcal{S}$, $|A| = (1+\theta)|\mathcal{S}|$ with equal probability. Let $E$ denote the random event that the set $\widehat{\mathcal{S}}$ is a stopping set. We can bound the probability of this random event as follows.

$$\begin{aligned}
\mathbb{P}\{E\} &\overset{(a)}{=} \sum_G \mathbb{P}(G_{k+1} = G)\mathbb{P}(E|G_{k+1} = G) \\
&\overset{(b)}{=} \sum_G \mathbb{P}(G_{k+1} = G)\frac{n_1(G)}{n_2(G)} \\
&\overset{(c)}{\geq} \sum_G \mathbb{P}(G_{k+1} = G)\frac{n_1(G)}{\exp\{[H_e(\theta) + \log_e(2)(1-\theta)]|\mathcal{S}|\}} \\
&= \exp\{-|\mathcal{S}|H_e(\theta) - (\log_e 2)(1-\theta)|\mathcal{S}|\}\,\mathbb{E}\left[\mathfrak{B}(\mathcal{S},\theta)\right] \qquad (8)
\end{aligned}$$

In the above inequality, (a) follows from the total probability theorem, where $G$ denotes the realization of the random graph $G_{k+1}$; (b) follows from a simple counting argument on the discrete probability space where $n_1(G)$ denote the number of stopping sets $A$ in $G$ with $\pi_{k+1}(A) = \mathcal{S}$, $|A| = (1+\theta)|\mathcal{S}|$, $n_2(G)$ denote the number of variable node sets $A$ in $G$ with $\pi_{k+1}(A) = \mathcal{S}$, $|A| = (1+\theta)|\mathcal{S}|$; (c) follows from the fact that the number of variable node set $A$, $\pi_{k+1}(A) = \mathcal{S}$ with cardinality $(1+\theta)|\mathcal{S}|$ is

$$\binom{|\mathcal{S}|}{\theta|\mathcal{S}|}2^{(1-\theta)|\mathcal{S}|} \qquad (9)$$

and the Stirling approximation [12, page 284]

$$\frac{2^{nH_2(k/n)}}{n+1} \leq \binom{n}{k} \leq 2^{nH_2(k/n)} \qquad (10)$$

Given the variable node set $\widehat{\mathcal{S}}$, let $\mathcal{S}_1$ denote the set of variable nodes $u$ in $\mathcal{S}$ with $\pi^{-1}(u) \subset \widehat{\mathcal{S}}$. Let $\mathcal{S}_0 = \mathcal{S}\backslash\mathcal{S}_1$. It is clear that $|\mathcal{S}_1| = \theta|\mathcal{S}|,\ |\mathcal{S}_0| = (1-\theta)|\mathcal{S}|$. Let the set $\mathcal{C}$ denote the set of check nodes $c$ such that $c$ is in $G_k$, and $c$ is connected to $\mathcal{S}_1$ at most once. We further consider an arbitrary chosen check node set $\widehat{\mathcal{C}}$ in $G_{k+1}$ such that $\pi_{k+1}(\widehat{\mathcal{C}}) = \mathcal{C}$, $|\widehat{\mathcal{C}}| = |\mathcal{C}|$. It is clear that for each pair of check nodes $c_1, c_2 \in \pi_{k+1}^{-1}(c)$ with $c \in \mathcal{C}$, exactly one of $c_1, c_2$ is in $\widehat{\mathcal{C}}$. By the hypothesis,

$$|\widehat{\mathcal{C}}| \geq (\beta - \gamma\theta)|\mathcal{S}| \qquad (11)$$

Let $m_j$ denote the number of check nodes $c \in \widehat{\mathcal{C}}$ such that $\pi_k(c)$ is not connected to any variable node in $\mathcal{S}_1$, $\pi_k(c)$ has degree $j$ in the subgraph induced by variable node set $\mathcal{S}$ and its neighboring check node set $N(\mathcal{S})$. Let $n_j$ denote the number of check nodes $c \in \widehat{\mathcal{C}}$ such that $\pi_k(c)$ is connected to exactly one variable node in $\mathcal{S}_1$, $\pi_k(c)$ has degree $j$ in the subgraph induced by variable node set $\mathcal{S}$ and its neighboring check node set $N(\mathcal{S})$.

The probability that the set $\widehat{\mathcal{S}}$ is a stopping set can be upper bounded as follows.

$$\begin{aligned}
\mathbb{P}\{E\} &\overset{(a)}{\leq} \prod_j \left\{1 - j\left(\frac{1}{2}\right)^j\right\}^{m_j} \prod_j \left\{1 - \left(\frac{1}{2}\right)^{j-1}\right\}^{n_j} \\
&= \exp\left\{\sum_j m_j \log_e \left[1 - j\left(\frac{1}{2}\right)^j\right]\right\} \exp\left\{\sum_j n_j \log_e \left[1 - \left(\frac{1}{2}\right)^{j-1}\right]\right\} \\
&\overset{(b)}{\leq} \exp\left\{(-1)\sum_j m_j j\left(\frac{1}{2}\right)^j\right\} \exp\left\{(-1)\sum_j n_j \left(\frac{1}{2}\right)^{j-1}\right\} \\
&\overset{(c)}{\leq} \exp\left\{(-2)\sum_j m_j \left(\frac{1}{2}\right)^j\right\} \exp\left\{(-2)\sum_j n_j \left(\frac{1}{2}\right)^j\right\} \\
&\overset{(d)}{\leq} \exp\left\{(-2)\left(\sum_j m_j + n_j\right)\left(\frac{1}{2}\right)^{(\beta\bar{d}/(\beta-\gamma\theta))}\right\} \\
&\overset{(e)}{\leq} \exp\left\{(-2)(\beta-\gamma\theta)|\mathcal{S}|\left(\frac{1}{2}\right)^{\beta\bar{d}/(\beta-\gamma\theta)}\right\}
\end{aligned} \tag{12}$$

The right hand side of (a) is the probability that the check node set $\widehat{\mathcal{C}}$ does not contain any check node that is connected to $\widehat{\mathcal{S}}$ exactly once. The inequality (b) follows from the inequality $\log_e(1-x) \leq -x$, for $0 < x < 1$. The inequality (c) follows from the fact that the degree $j$ is always greater than or equal to 2. The inequality (d) follows from the convexity of exponential functions and that the average of degree $j$ is less than $\beta\bar{d}/(\beta-\gamma\theta)$. The inequality (e) follows from the Ineq. 11.

The Theorem follows from the Ineq. (8), and (12).

## VII. Numerical Results

In this Section, simulation results for the proposed 2-lift based LDPC codes are presented. We investigate both codes with algebraically constructed protographs and codes with protographs obtained by computational searching.

We construct a (3,4) regular code with blocklength 992 bits, rate $0.27$. The protograph is chosen as one LDPC code previously reported in [13]. The protograph is a (3,4)-regular code with blocklength 124, dimension 33, minimal distance $24$. We find that the protograph does not contain small stopping set by exhaustive searching. The parity-check matrix corresponding to the protograph is as follows:

$$H = \begin{bmatrix} I_1 & I_2 & I_4 & I_8 \\ I_5 & I_{10} & I_{20} & I_9 \\ I_{25} & I_{19} & I_7 & I_{14} \end{bmatrix}$$

where, $I_k$ represents $k$-times cyclically right-shifted identity matrices of size $31 \times 31$.

We depict the error probability performance of the above $2$-lift based code in Figs. 3, and 4. The code is compared with a random $(3,4)$ regular LDPC code. The channels are BECs with various channel parameters $\xi$. The error probabilities of the proposed code are represented by solid curves, while the error probabilities of the random $(3,4)$ regular code are represented by dashed curves. The two codes have almost identical error probability performance.

We also construct codes with protographs obtained by computational searching. We show the parity-check matrix of a protograph in Fig. 5. The protograph does not contain low-weight stopping sets with weak graph expansion properties. The protograph contains $32$ variable nodes and $16$ check nodes. The degree distributions of the protograph is chosen to approximate the degree distributions in Eqs. 13, 14, which are obtained by density evolution [14]. The blocklength of the resulting code is $8192$. The rate is $0.5$.

$$\lambda(x) = 0.1896x + 0.4806x^2 + 0.1032x^{11} + 0.2266x^{12} \tag{13}$$

$$\rho(x) = 0.8693x^6 + 0.1307x^7 \tag{14}$$

We depict the error probability performance of the above $2$-lift based code in Figs. 6, and 7. The code is compared with a random irregular code with degree distributions in Eqs. 13, 14 and a random $(3,6)$ regular LDPC code. The channels are BECs with various channel parameters $\xi$. The $2$-lift base code has a waterfall position close to that of the random irregular code. However, the proposed code does not exhibit an error floor as the random irregular code. The $2$-lift base code outperforms the random $(3,6)$ regular code in terms of error probability performance, because the code has a better waterfall position.

## VIII. Conclusion

This paper proposes a new construction schemes for LDPC codes based on random $2$-lift. We present an analysis for low-weight stopping set distributions. Based on this analysis, we propose design criteria for the resulting codes to have low error floors. Numerical results show that the proposed codes outperform the previously constructed codes in terms of error probability performance.

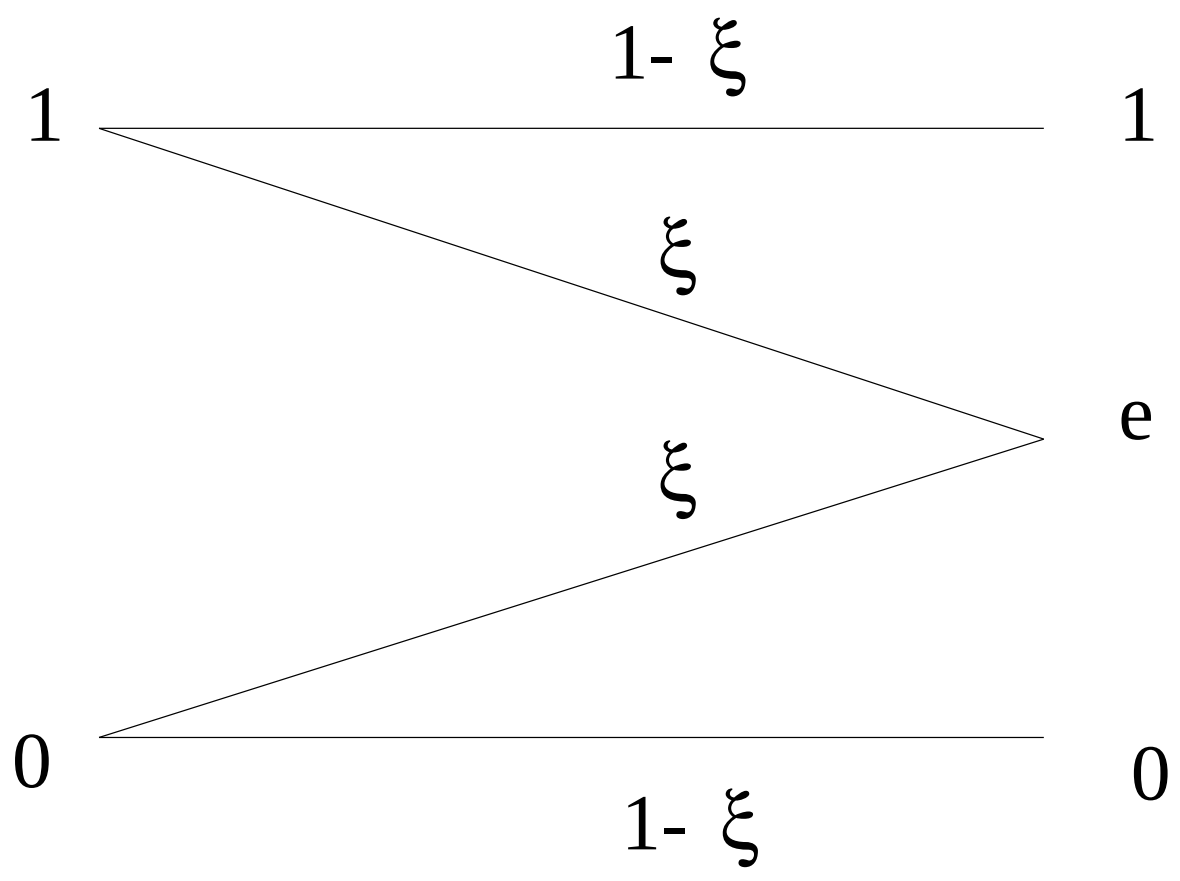

Fig. 1. Binary erasure channel

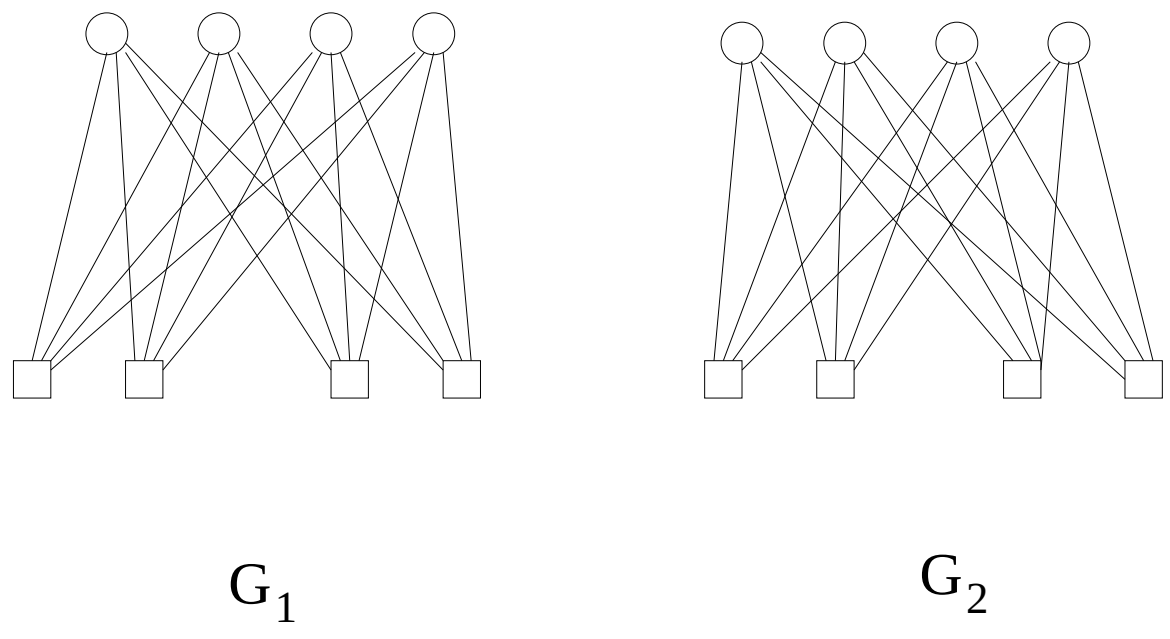

(a) Two copies of the base graph

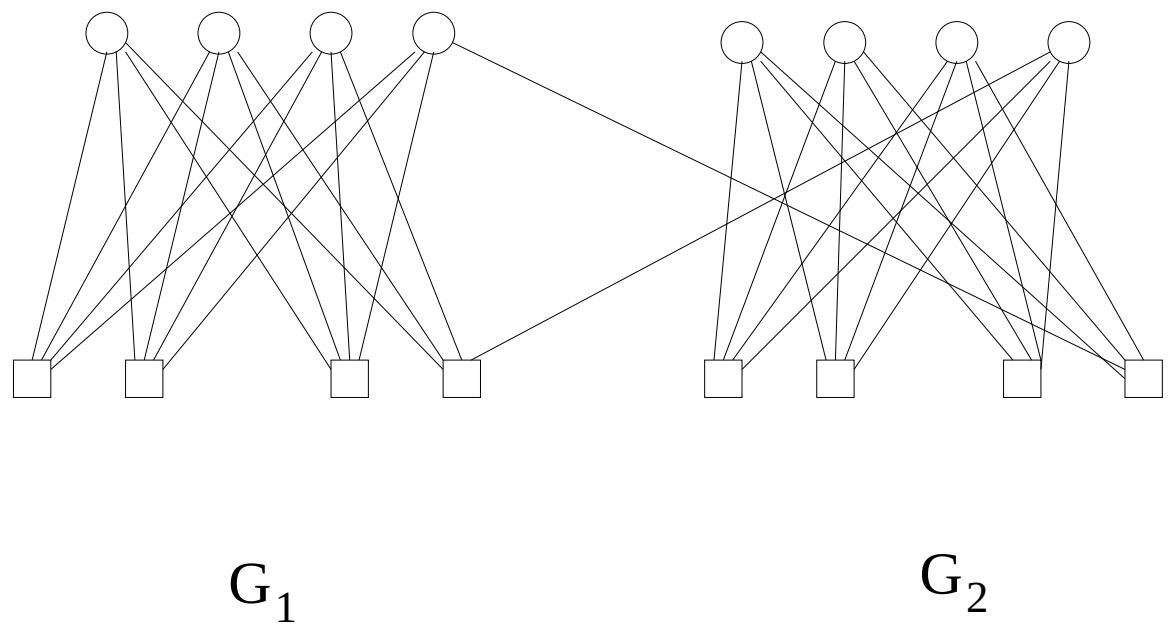

(b) A 2-lift with one edge permutation

Fig. 2. Examples of graph lift

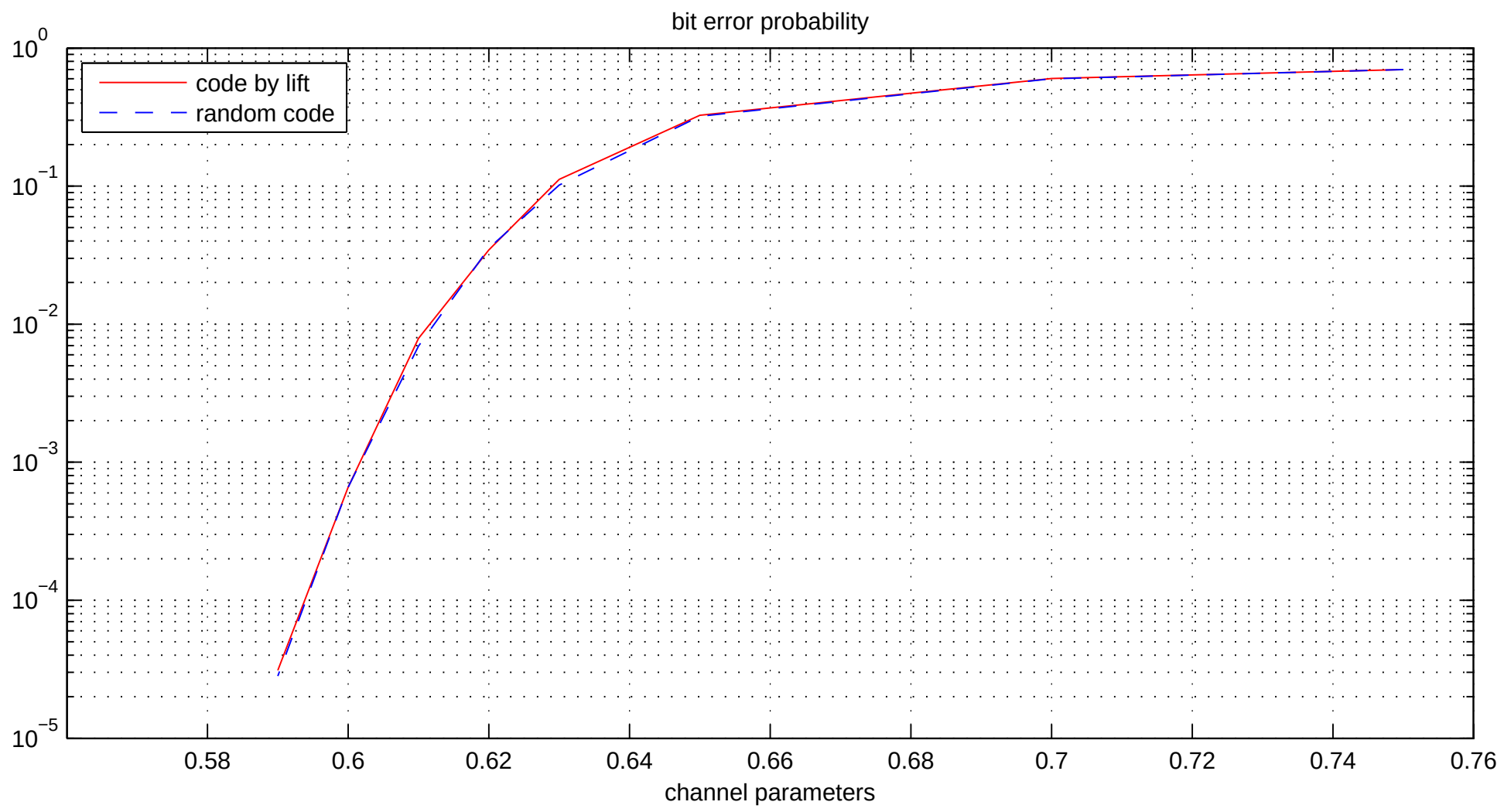

Fig. 3. Bit error probability of a 2-lift based code with an algebraically constructed protograph. The channels are BECs with various channel parameters.

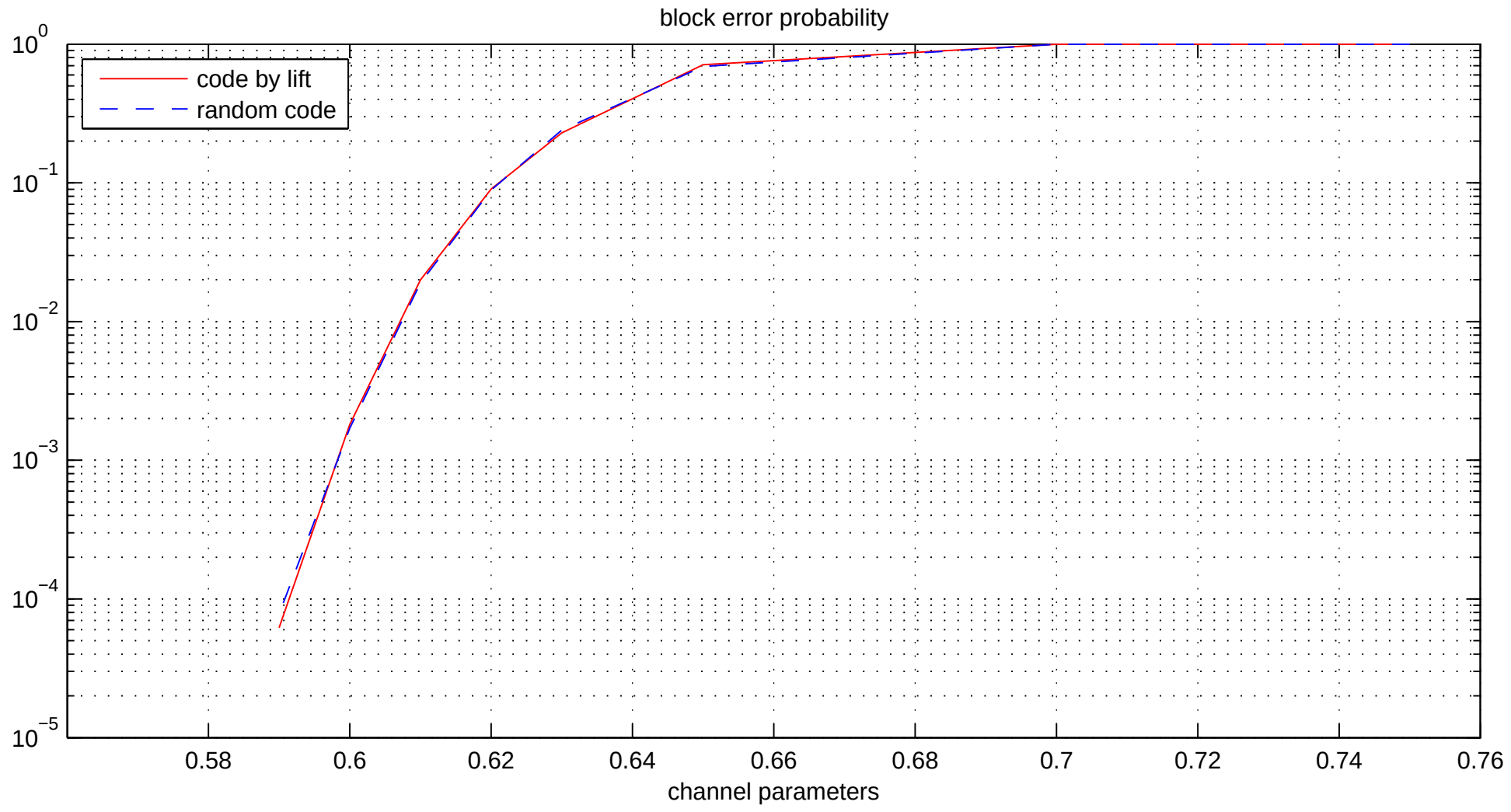

Fig. 4. Block error probability of a 2-lift based code with an algebraically constructed protograph. The channels are BECs with various channel parameters.

$$
\begin{bmatrix}
0 & 0 & 0 & 0 & 0 & 0 & 0 & 0 & 0 & 0 & 1 & 0 & 0 & 1 & 0 & 0 \\
0 & 1 & 0 & 1 & 0 & 0 & 0 & 0 & 0 & 0 & 0 & 0 & 0 & 0 & 0 & 0 \\
0 & 0 & 0 & 1 & 0 & 0 & 0 & 0 & 0 & 0 & 0 & 0 & 0 & 0 & 0 & 1 \\
0 & 0 & 0 & 0 & 1 & 0 & 0 & 0 & 0 & 0 & 0 & 0 & 1 & 0 & 0 & 0 \\
1 & 0 & 0 & 0 & 0 & 0 & 0 & 0 & 1 & 0 & 0 & 0 & 0 & 0 & 0 & 0 \\
0 & 0 & 0 & 0 & 0 & 0 & 1 & 0 & 0 & 0 & 0 & 1 & 0 & 0 & 0 & 0 \\
0 & 0 & 0 & 0 & 1 & 0 & 0 & 1 & 0 & 0 & 0 & 0 & 0 & 0 & 0 & 0 \\
0 & 0 & 0 & 0 & 0 & 1 & 0 & 0 & 0 & 0 & 0 & 0 & 0 & 0 & 0 & 1 \\
0 & 0 & 0 & 0 & 0 & 0 & 0 & 1 & 0 & 0 & 1 & 0 & 0 & 0 & 0 & 0 \\
0 & 0 & 0 & 0 & 0 & 0 & 0 & 0 & 0 & 0 & 0 & 0 & 1 & 1 & 0 & 0 \\
0 & 0 & 0 & 1 & 0 & 0 & 0 & 0 & 1 & 0 & 0 & 0 & 0 & 0 & 0 & 1 \\
0 & 0 & 1 & 0 & 0 & 0 & 1 & 0 & 0 & 0 & 0 & 1 & 0 & 0 & 0 & 0 \\
0 & 0 & 0 & 0 & 1 & 0 & 0 & 1 & 0 & 0 & 0 & 0 & 0 & 1 & 0 & 0 \\
0 & 0 & 0 & 0 & 0 & 0 & 0 & 0 & 0 & 0 & 0 & 1 & 0 & 0 & 1 & 1 \\
1 & 0 & 0 & 0 & 0 & 1 & 1 & 0 & 0 & 0 & 0 & 0 & 0 & 0 & 0 & 0 \\
0 & 0 & 1 & 0 & 0 & 0 & 0 & 0 & 1 & 0 & 0 & 0 & 1 & 0 & 0 & 0 \\
0 & 1 & 0 & 0 & 0 & 0 & 0 & 0 & 0 & 0 & 0 & 1 & 0 & 0 & 0 & 1 \\
1 & 0 & 0 & 0 & 0 & 0 & 0 & 1 & 0 & 0 & 0 & 0 & 1 & 0 & 0 & 0 \\
0 & 0 & 0 & 0 & 0 & 0 & 0 & 0 & 1 & 0 & 0 & 0 & 1 & 0 & 1 & 0 \\
1 & 0 & 0 & 0 & 0 & 0 & 1 & 0 & 0 & 0 & 0 & 0 & 0 & 0 & 0 & 1 \\
0 & 0 & 0 & 0 & 1 & 0 & 1 & 0 & 0 & 0 & 0 & 1 & 0 & 0 & 0 & 0 \\
0 & 0 & 1 & 0 & 0 & 0 & 1 & 0 & 0 & 0 & 0 & 1 & 0 & 0 & 0 & 0 \\
0 & 0 & 0 & 1 & 0 & 0 & 0 & 0 & 0 & 1 & 0 & 0 & 0 & 0 & 1 & 0 \\
0 & 1 & 0 & 0 & 0 & 0 & 0 & 0 & 1 & 0 & 0 & 1 & 0 & 0 & 0 & 0 \\
1 & 0 & 0 & 0 & 0 & 0 & 1 & 0 & 0 & 0 & 0 & 0 & 0 & 0 & 0 & 1 \\
0 & 0 & 0 & 1 & 0 & 0 & 1 & 0 & 0 & 0 & 0 & 0 & 0 & 0 & 1 & 0 \\
0 & 1 & 1 & 0 & 0 & 0 & 0 & 0 & 0 & 0 & 0 & 0 & 1 & 0 & 0 & 0 \\
0 & 1 & 0 & 0 & 1 & 0 & 0 & 0 & 0 & 0 & 0 & 0 & 1 & 0 & 0 & 0 \\
1 & 1 & 1 & 0 & 1 & 1 & 0 & 1 & 0 & 1 & 1 & 1 & 1 & 1 & 1 & 1 \\
1 & 0 & 1 & 1 & 1 & 1 & 0 & 1 & 1 & 1 & 1 & 1 & 0 & 1 & 1 & 0 \\
1 & 1 & 1 & 1 & 1 & 1 & 0 & 1 & 1 & 1 & 1 & 0 & 0 & 1 & 1 & 0 \\
0 & 1 & 1 & 1 & 0 & 1 & 0 & 1 & 1 & 1 & 1 & 0 & 1 & 1 & 1 & 1
\end{bmatrix}
$$

Fig. 5. The parity-check matrix corresponds to a protograph obtained by computational searching. The rows correspond to variable nodes. The columns correspond to parity-check equations.

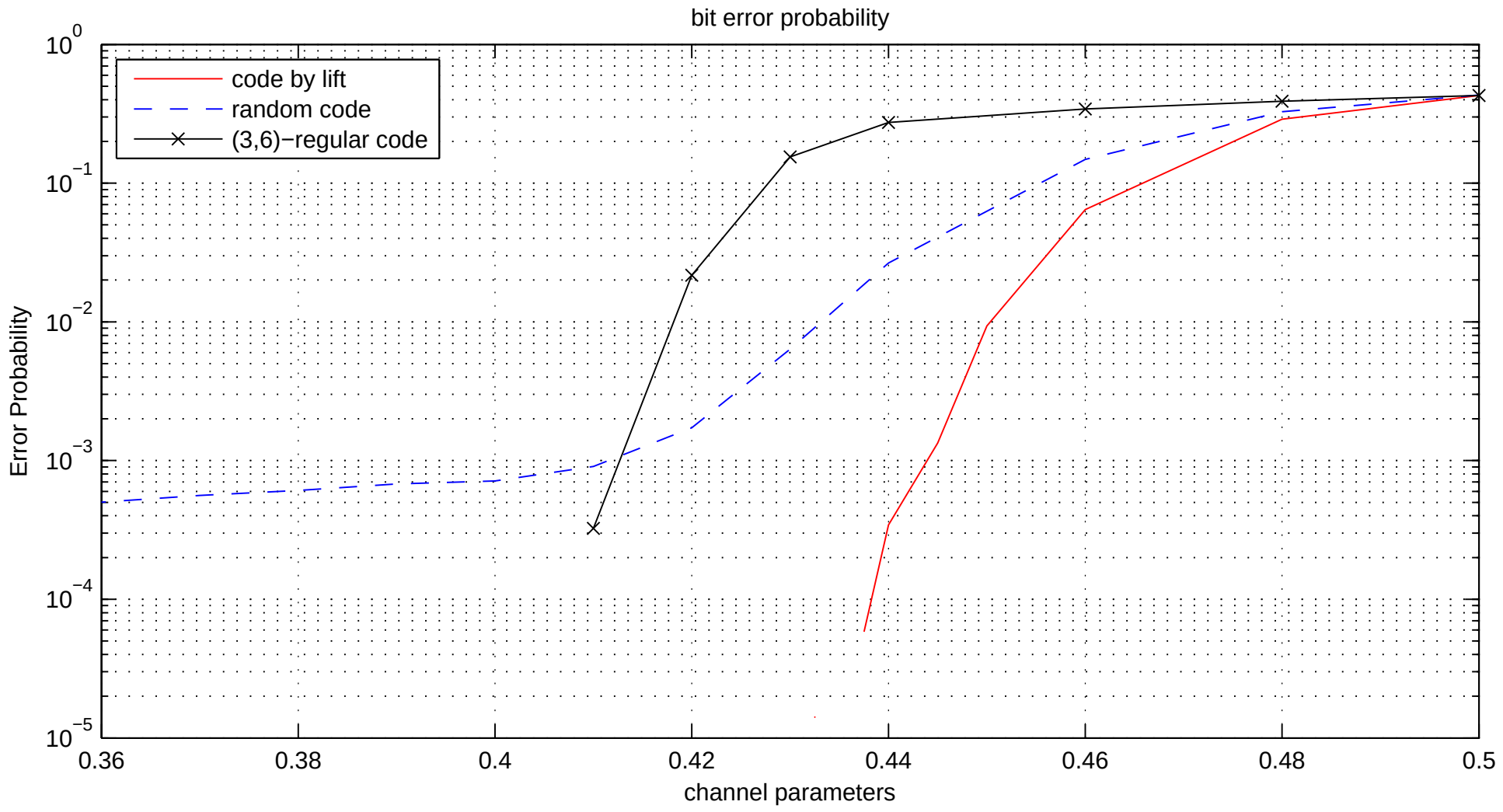

Fig. 6. Bit error probability of a 2-lift base code over BECs.

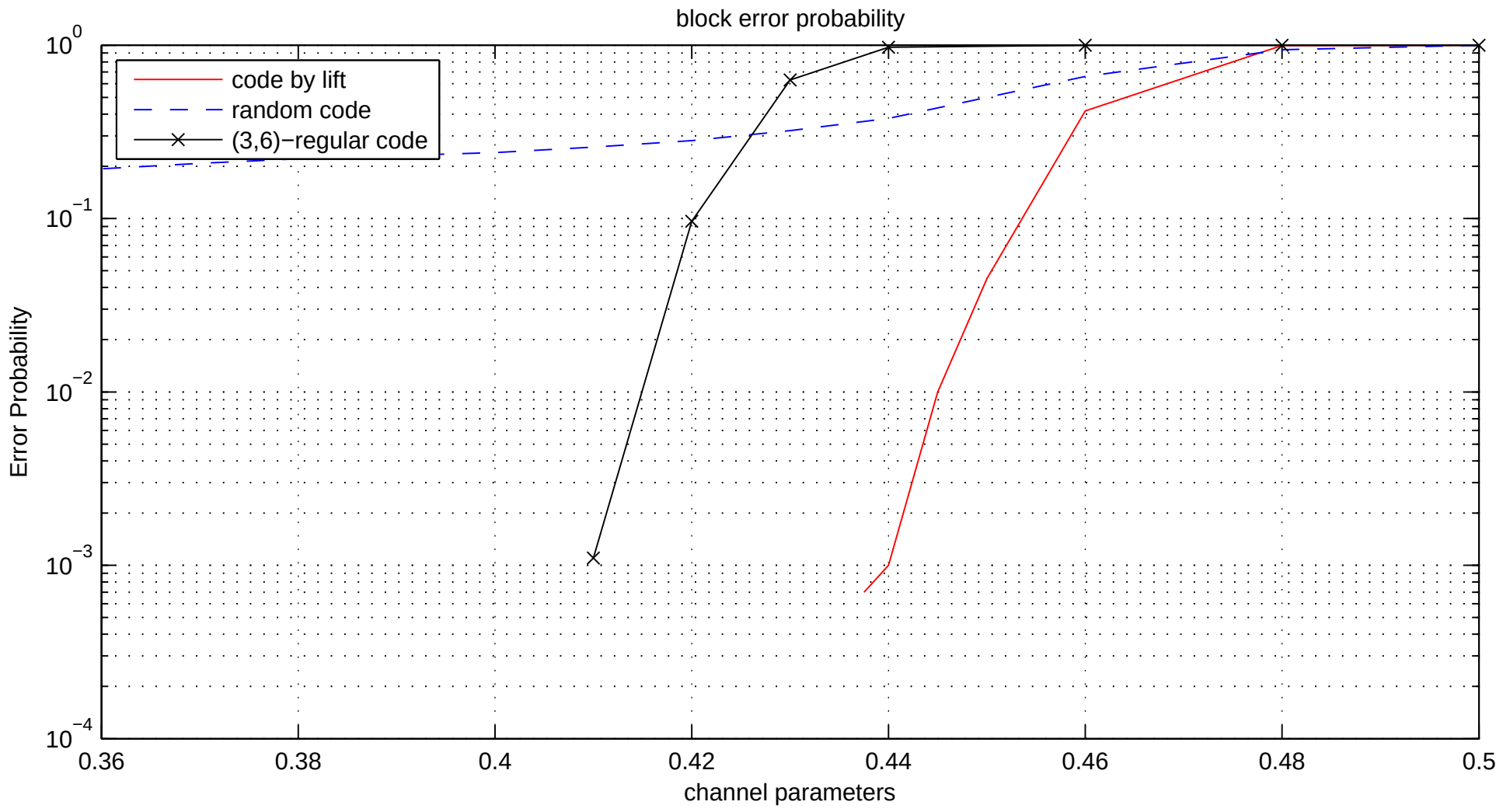

Fig. 7. Block error probability of a 2-lift base code over BECs.